# BioInfoBase : A Bioinformatics Resourceome

Kadkhodaei S[1*], Barantalab F[2], Taheri S[3], Foroughi M[4], Hashemi FG[1], Shabanimofrad MR[3], Hosseinimonfared H[5], Rezaei MA[5], Ranjbarfard A[6], Sahebi M[1], Azizi P[1], Dadar M[7], Abiri R[4], Harighi MF[3], Kalhori N[4], Etemadi MR[2], Baradaran A[8], Danaee M[9], Zare I[10], Ghafarpour A[10], Azhdari Z[4], Rajabi Memari H[11], Safavi V[12], Tajabadi N[13], Faruku Bande[14]

[1] Institute of Tropical Agriculture, Universiti Putra Malaysia, Serdang 43400, Malaysia
[2] Faculty of Medicine, Universiti Putra Malaysia, Serdang 43400, Malaysia
[3] Faculty of Agrobiotechnology, Universiti Putra Malaysia, Serdang 43400, Malaysia
[4] Faculty of Biotechnology and Biomolecular Sciences, Universiti Putra Malaysia, Serdang 43400, Malaysia
[5] Faculty of Veterinary Medicine, Universiti Putra Malaysia, Serdang 43400, Malaysia
[6] Center of Plant Biotechnology (CPB), National University of Malaysia, 43600 Selangor, Malaysia
[7] Center of Biotechnology and Biology Research, Shahid Chamran University, Ahvaz, Iran
[8] Diamantina Institute, University of Queensland 36992, Australia
[9] Faculty of Medicine, University of Malaya, Kuala Lumpur 59990, Malaysia
[10] Department of Cellular and Molecular Biology, Semnan University, Semnan, Iran
[11] SynHiTech, Thornhill, Ontario L3T 0C7, Canada
[12] Institute of Biology, Dahlem Centre of Plant Sciences, Free University Berlin, Albrecht-Thaer-Weg 6, D-14195 Berlin, Germany
[13] Animal Science Research Institute, Dehghanvila 1, Karaj, Iran
[14] Ministry of Animal Health and Fisheries Development, PMB 2109 Usman Faruk Secretariat Sokoto, Nigeria

* Corresponding author: admin@bioinfobase.info

**Abstract**

Over the past decade there has been a significant growth in bioinformatics databases, tools and resources. Although, bioinformatics is becoming more specific, increasing the number of bioinformatics-wares has made it difficult for researchers to find the most appropriate databases, tools or methods which match their needs. Our coordinated effort has been planned to establish a reference website in Bioinformatics as a public repository of tools, databases, directories and resources annotated with contextual information and organized by functional relevance. Within the first phase of BioInfoBase development, 22 experts in different fields of molecular biology contributed and more than 2500 records were registered, which are increasing daily. For each record submitted to the database of website almost all related data (40 features) has been extracted. These include information from the biological category and subcategory to the scientific article and developer information. Searching the query keyword(s) returns links containing the entered keyword(s) found within the different features of the records with more weights on the title, abstract and application fields. The search results simply provide the users with the most informative features of the records to select the most suitable ones. The usefulness of the returned results is ranked according to the matching score based on the Term Frequency-Inverse Document Frequency (TF-IDF) methods. Therefore, this search engine will screen a comprehensive index of bioinformatics tools, databases and resources and provide the best suited records (links) to the researchers' need. The BioInfoBase resource is available at [www.bioinfobase.info](www.bioinfobase.info).

**Introduction**

The rapid growth rate of bioinformatics during the last decade has resulted in developing thousands of tools, databases, and resources. Because of this fast growth, it is inconceivable for bioinformaticians and molecular biologist to use all of these databases, tools, and resources. Additionally, understanding all capabilities of these resources could assist these researchers, however, the abundance of them making it impossible. On the other hand, due to the large increasing number of these resources it is difficult for bioinformatics researchers to find the most suited methods, tools and databases in the area and/or even within their own specialty field. Furthermore, biologist users approaching the field need a comprehensive directory of bioinformatics algorithms, databases, tools, and literature explained with information about their framework and appropriate application. In response to all these needs and problems, the BioInfoBase was developed.

There are a few resources in this regards including BioNetBook (project discontinued) (1), Bioinformatics Link Directory ([http://bioinformatics.ca/links_directory](http://bioinformatics.ca/links_directory)) (2–5), Online Bioinformatics Resources Collection: OBRC ([http://www.hsls.pitt.edu/obrc/](http://www.hsls.pitt.edu/obrc/)) (6), Startbioinfo ([http://shodhaka.com/startbioinfo](http://shodhaka.com/startbioinfo)) (7), Weblab ([http://weblab.cbi.pku.edu.cn](http://weblab.cbi.pku.edu.cn)) (8), Softberry ([http://www.softberry.com](http://www.softberry.com)) (9), OmicTools ([http://omictools.com](http://omictools.com)) (10), The

Elements of Bioinformatics (http://elements.eaglegenomics.com) (11), and Phylogeny tools (http://evolution.genetics.washington.edu/phylip/software.html) (12). However, in most cases the user friendliness, comprehensiveness and access to the most relevant tools and/or databases are on questions.

In scientific searching, the researcher provides the search engine with the keyword(s) about a subject which the user is trying to collect research information. The researcher attempts to find the documents which together will arrange for the sought after information. The most basic and key rule for search technology developments is to make the users closer to results. This is always about faster and more accurate results and the direction where search engines head. There are many bioinformatics related subjects which the numbers are increasingly growing. Furthermore, for the same function (e.g. signal peptide prediction) the searcher may find many software and/or databases among which the "best" or the "closest" is an open question. On the other hand, sometimes the searcher is not aware of what exactly s/he is looking for.

In this paper, we report on development of a comprehensive databank of bioinformatics tools, databases, directories and resources (annotated with contextual information and organized by functional relevance) as well as the specific search engine. The proposed database and searching method is significant because it facilitates the access to the best suited records (links) to the researcher's need. The comprehensiveness of the database, user-friendliness of the interface as well as the efficiency of the specific search engine compared with the similar websites could make it popular as a "Bioinformatics Google" or "Bioinformatics Wikipedia".

**Materials and Methods**

*Data mining, preprocessing and management*

Figure 1 illustrates the workflow of BioInfoBase design. All relevant records were collected over the past 5 years particularly from high-quality publications and leading journals of *Nucleic Acids Research* and *Bioinformatics*. Two levels of classifications were applied to the records based on; (i) the nature of the record (Software, Database, Directory, Resource and Miscellaneous) and (ii) the Omics relevance of the record (Genomics, Transcriptomics, Proteomics, Metabolomics, etc.).

For each record submitted to the website's database almost all related data (40 features) has been extracted (Table 1). These include information from the biological category and subcategory to the paper and developer information. A platform was designed to manage (submission, editing, maintenance and monitor) the records (Figure 2).

The records could be submitted either directly by a registrar (who might be the developer of the program) or through the pending list which is already provided by BioInfoBase administrators. The pending list includes a number of records which their information needs to be extracted and submitted to the BioInfoBase database. Wiki style as a quick collaboration technology on the web was used to virtually enable everyone in editing the pages without difficulty. However, the information managed and finalized by the website admins.

### *Databasing and interface design*

The BioInfoBase network management platform was designed and developed using PHP programming language, CSS3 (13) and HTML5 (14) technologies, and MySQL database management system (http://www.mysql.com) under YII FrameWork (http://www.yiiframework.com) (15). The YII FrameWork was used as an open source and fast web server which speed, security, flexibility and compliance are all its characteristics comparing to other competitors. The database is hosted on an Apache server as it collaborates well with PHP and MySQL technologies.

### *Search engine development*

The method of choice to find the keywords was the one that was suggested and tested by Rose et al. (16). The Rapid Automatic Keyword Extraction (RAKE) was developed which is based on the extraction of keywords by deconstructing the text into a set of the candidate terms (keywords). The weights of the terms were obtained by using Term Frequency-Inverse Document Frequency (TF-IDF) methods. Weight of a term (keyword) was defined by the frequency of the term appearance in a document (record). This is denoted as the TF (Term Frequency), and the number of documents that comprise the specific term is denoted as the IDF (Inverse Document Frequency); and is called the TF-IDF weight (17). The following formula describes the basic form of TF-IDF:

$$W_{i,j} = TF_{i,j} * IDF_i \qquad \text{Equation 1}$$

were;

$W_{i,j}$: Weight of term *i* in document *j*

$TF_{i,j}$: Frequency for term *i* in document *j*; which is the local weight for term *i*

$IDF_i$: Inverse document frequency of term *i*; which is the global weight for term *i*.

$$IDF_i = \text{Log}\ \frac{D}{d_i} \qquad \text{Equation 2}$$

were;

D: Number of documents in the database

$d_i$: Number of documents containing term *i*.

If the term is not in the database, this will result in an error of division-by-zero. This problem was resolved by modifying the formula to:

$$IDF_i = \log \frac{D}{1+d_i} \qquad \text{Equation 3}$$

Since the document length is also an affecting term weighting factor, a normalization factor was used to improve the inconsistencies in this regards. In our case, the normalization was based on the following equation proposed by (18):

$$N_{ij} = \frac{TF_{ij}}{\max(TF_j)} \qquad \text{Equation 4}$$

were;

$N_{ij}$: Normalization factor of term *i* in document *j*
$TF_{ij}$: Frequency for term *i* in document *j*
$\max(TF_j)$: Maximum *TF*s in document *j*

The final term weighting formula using TF-IDF model considering normalization factor was:

$$W_{i,j} = TF_{i,j} * IDF_i * N_{ij} \qquad \text{Equation 5}$$

were;

$W_{i,j}$: Weight of term *i* in document *j*
$TF_{i,j}$: Frequency for term *i* in document *j*
$IDF_i$: Inverse document frequency of term *i*
$N_{ij}$: Normalization factor of term *i* in document *j*

*BioInfoBase mobile application development*

The Android devices and smartphones comprise the potential of increasing apps productivity without the effects of spatial constraints on users. We developed BioInfoBase Android mobile application to simplify the access to bioinformatics tools and databases through BioInfoBase interface. Using this app the user will be navigated to launch the mobile browser and the search box of the BioInfoBase. The BioInfoBase Android app was built using the Java programming language and with the standard API provided for Android development. BioInfoBase for Android was developed using version 4.3.0.v20130605-2000 of the Eclipse (Kepler Release) Integrated Environment. The Android version of the BioInfoBase app is available for Android version 4.1 onward and is compatible with Android devices such as smartphones and tablet PCs. Later versions of additional app will be developed updated with novel functionalities in the near future. The BioInfoBase Android app is freely available on http://www.bioinfobase.info/.

**Results and Discussion**

*The structure of BioInfoBase*

The BioInfoBase database is managed by a MySQL interactive database that offers the back-end for user queries and facilitates retrieval of the data necessary for the site's analysis tools. The method of choice (16) to develop the BioInfoBase search engine focuses on the automatically extraction of terms from documents to generate documents' summary features and/or to suggest keywords for an indexer (19).

The motivation in developing RAKE (Rapid Automatic Keyword Extraction) was to generate an extremely efficient keyword extraction method, which runs on separable documents to facilitate the application to dynamic databases. Consequently, it is simply functional to new fields, and works well on various documents types, especially those that do not follow particular grammar conventions. RAKE is based on the observation that terms frequently comprise multiple words but infrequently contain stop words or standard punctuation, including the function words *the*, *and*, *of*, and or other words with least vocabulary meaning. RAKE uses phrase delimiters and stop words to partition the record content into candidate terms, which are sequences of substantive words as they take place in the record's text. Because of the importance of some keywords over the others, we applied basic term weighting technique to demonstrate how a keyword is important in the documents for a collection or text.

Keyword extraction on a document in RAKE is initiated by fractionating the text into a set of substantive terms. Initially, the record text is fragmented into a collection of terms by the indicated term delimiters. This collection is then split into sequences of adjacent words at stop word and phrase delimiters positions. The same position in the text will be allocated to the terms within a sequence and it will be considered together as a candidate keyword. The weights of the keywords obtain by using TF-IDF methods.

The BioInfoBase currently hosts information for nonredundant records of bioinformatics tools and databases from different areas of molecular biology and a wide range of genomics, transcriptomics, proteomics, metabolomics, etc. The BioInfoBase database contains the results of three subsequent computational analyses performed on the records using the in-house search engine. The analysis processes include "Terms Extraction", "Calculation of Global Weight" and "Term Weighting", respectively.

In the search result interface, we aimed at: (i) providing rapid access to the most suited scientific databases and tools (bioinformatics resources), (ii) serving an inclusive range of users including beginners having little experience even not familiar with the fields. Searching the query keyword(s) returns a comprehensive list of the links containing the entered keyword(s) found within the different features of the records. The result page of the website offers several different weblink representations sorted by the matching score to meet the user needs. The search results simply provide the users with the most informative features of the records to enable them in selection of the most suitable ones (Figure 3). Therefore, the resulted records in browse result page of the BioInfoBase are accompanied by "Matching Score" and descriptions of the record's "Applications", "Categories", "Google Scholar Citation", "Features" and "Abstract" of the related publication. The detail information of the records is provided in "more" section, presenting a table containing all information associated with the weblinks (Figure 4). The table then presents the links to the record homepage, tutorial page and article page.

The usefulness and ranking of the records could be assessed based on matching score (term weighting method), google scholar citation and user ranking. The usefulness of the returned results is basically ranked by the matching score according to the TF-IDF method. Therefore, this search engine will screen a comprehensive index of bioinformatics tools, databases and resources and provides a list of the best suited records (links) to the researchers' needs. The search engine algorithm was programed in a way to consider more weights on the "Name", "Abstract", "Application" and "Features" fields in the records. On the other hand, validity of the records could be assessed by google scholar citation index and/or user ranking. The ability of ranking of the resulted records based on these two features will be provided in the next version.

In BioInfoBase, each record is maintained regularly and independently but many of them still follow the database update schedule. As a wiki-style collaboration project managed by bioinformaticians, everyone in molecular biology or in particular, bioinformatics community will have the opportunity to register and help in enrichment of this bioinformatics resourceome. The second phase of BioInfoBase is mainly focusing on development of an Intelligent Semantic Search Engine, providing a short introduction and/or tutorial film for each record, and keeping the database of website updated constantly.

*The statistics of BioInfoBase*

Within the first phase of BioInfoBase development (2010-2015), 22 experts in varying fields of molecular biology contributed to register more than 2500 nonredundant records, which are increasing daily. All relevant records were collected over the past 10 years particularly from high-quality publications and leading journals of *Nucleic Acids Research* and *Bioinformatics.* All records have been evaluated from 40 different views and all practical information for each submitted record has been extracted. These features contain information from the application and biological category to the papers and developer information. By searching the query keyword(s), the website will provide a comprehensive list of tools, databases, and resources based on their TF-IDF score.

For each record the researchers could find the most informative features such as "Matching Score" and descriptions of the record's "Applications", "Categories", "Google Scholar Citation", "Features" and "Abstract".

Furthermore, to evaluate the usage and quality of records we monitor the database on a regular basis and warn the record's developer(s) regarding the possible errors if any. Moreover, the popularity of records is monitored through usage statistics for annual report.

Last but not the least, in order to assess the world bioinformatics trend, we performed a preliminary study on our database. As a brief overview on the world bioinformatics resources landscape according to the submitted records (Table 2, Figure S1):

- All records placed in four major categories of molecular biology as follows: Genomics (48.04%), Proteomics (16.24%), Transcriptomics (7.2%) and Metabolomics (5.52%).

- The quantity of softwares was almost twice more than databases.

- Approximately 40 percent of the recorded papers were published by 7 leading journals: Nucleic Acids Research (26.2%), Bioinformatics (5.52%), BMC Bioinformatics (2.12%), Biotechniques (2.08%), Nature Group (1.4%), Genome Research (1.24%) and Briefing in Bioinformatics (0.48%).

- About 70 percent of the records were open access. Additionally, the amount of on-line records was 3.5 times more than the stand-alone ones.

- The ten pioneer countries which generated more than half of the scientific productions in the field of Bioinformatics were: USA (23.28%), UK (7.88%), Germany (4.96%), Canada (4.76%), Japan (2.92%), France (3.84%), India (1.4%), Denmark (1.32%), Italy (1.32%) and China (1.2%).

- The following institutions had the largest portion in generating these publications: European Molecular Biology Laboratory-European Bioinformatics Institute (EMBL-EBI), Bioinformatics and Genomics Department of Agricultural-Food and Nutritional Science (AFNS), National Center for Biotechnology Information (NCBI), National Institute of Health (NIH), Swiss Bioinformatics Institute (SIB) and Institut national de la recherche agronomique (INRA).

- The total number of citations for the records having a published scientific article was 673,502 with an average of 544.9. This high number of citations demonstrates the significance and increasing development of bioinformatics in last decade.

This analysis is in process to generate a comprehensive illustration of bioinformatics resource landscape which will be released in the next phase of BioInfoBase development.

**Future Directions**

The second phase of BioInfoBase is mainly focusing on the following:

(1) Development of an Intelligent Semantic Search Engine (Ontology-Based Semantic Search) to improve the search engine performance;

Semantic search tries to improve accuracy of search by understanding searcher intent and the contextual meaning of terms as they appear in the searchable data space, whether within a closed system or on the Web, to produce more applicable results. Our effort is something like Google or Bing in Bioinformatics. However, instead of using ranking algorithms such as PageRank (in Google) to predict relevancy, semantic search utilizes semantics, or the science of meaning in language, to generate highly relevant search results rather than a list of loosely related records.

(2) Advanced search for a higher precision query; a customizable multiple keyword search allowing the user to restrict the query and search results to a subset or particular database fields.

(3) Keeping the website database updated constantly.

(4) Development of a comprehensive illustration of world bioinformatics resource landscape.

**Conclusion**

In this paper, we report on development of the most comprehensive database of bioinformatics tools, databases, directories and resources (annotated with contextual information and functional classification) as well as the specific search engine. The method of search is based on the terms weight obtained by using TF-IDF. The proposed database and searching method is significant because it facilitates the access to the best suited records (links) to the researcher's need. The comprehensiveness of the database besides the user-friendliness of the interface compared with similar websites could make it popular as a "Bioinformatics Google" or "Bioinformatics Wikipedia". The related website should be of interest to readers in the all areas of Molecular Biology including Genomics, Transcriptomics, Proteomics and Metabolomics.

In summary, BioInfoBase provides:

- Rapid access to the most suited bioinformatics databases and tools in different omics areas.
- Comprehensive list of resources released by numerous and various institutions
- User friendly search functionality to allow proper access to the independent resources

- Simple and efficient search result interface
- Regular and individual maintenance of the records
- Monitoring of the quality and usage of resources

We encourage users of the database to register at "http://bioinfobase.info/search_home.php?r=user/login" to access the detail information of the records. We also encourage authors and developers of bioinformatics tools and databases to register if they wish to add their programs to the BioInfoBase database following the publication acceptance.

**Availability**

All data hosted by the BioInfoBase database are freely available at http://www.bioinfobase.info. All pages have been tested under Firefox, Internet Explorer and Chrome browsers.

**Acknowledgements**

We thank A. Moosavifar for technical assistance and R. Beheshtifar for manuscript preparation.

# FIGURES

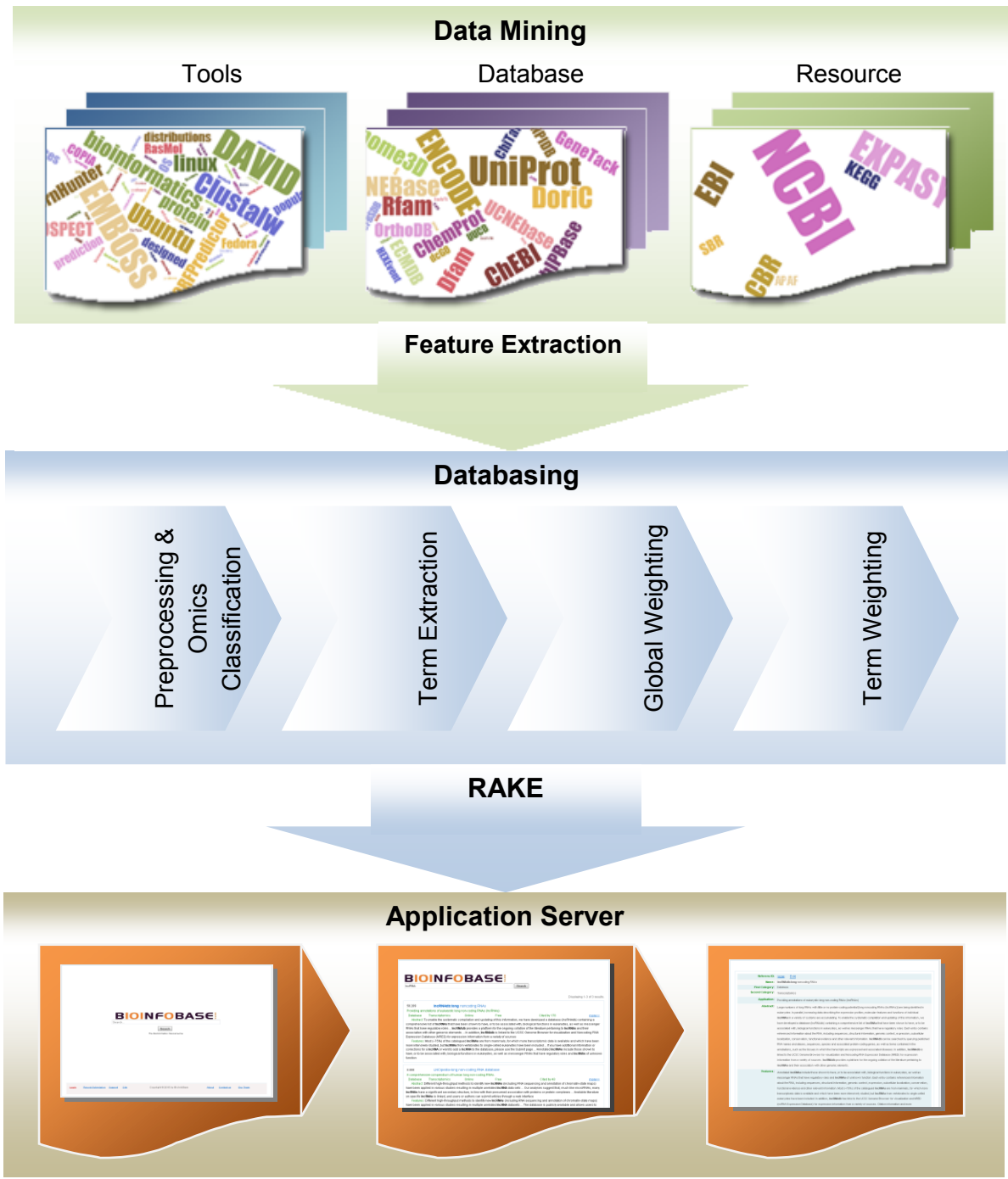

**Figure 1.** The schematic workflow of BioInfoBase design (Beta version). (1) Data mining: Applying the selection criteria, first all the relevant records were collected mainly from literatures and processed for feature extraction. (2) Databasing: The structured data tables used as input to MySQL (http://www.mysql.com/) to generate searchable data tables by end user. All data were processed for Rapid Automatic Keyword Extraction (RAKE) through four main steps; preprocessing, term extraction, global weighting and term weighting, respectively. (3) The web application and the interface were implemented to query the database.

**Figure 2.** The BioInfoBase control panel enables the admins to manage records (Record Submission), Monitor the quality and usage of records (Search Engine Log) and Registrars (Profile).

**Figure 3.** The search result page of the BioInfoBase (Beta version). The more informative features have been selected to be displayed in the search result page. These include "Application", "First Category", "Second Category", "Availability", "Scholar Citation", "Abstract" and "Main Features". The "more" will direct the user to the record detailed information.

**Figure 4.** The "more" item on the right-hand side of each record in result page allows users to discover detailed information of the record, which might help them in their work. If the user is the Record Registrar/Record Developer, s/he will be able to directly edit or update the record in this page.

**SUPPLEMENTARY FIGURES**

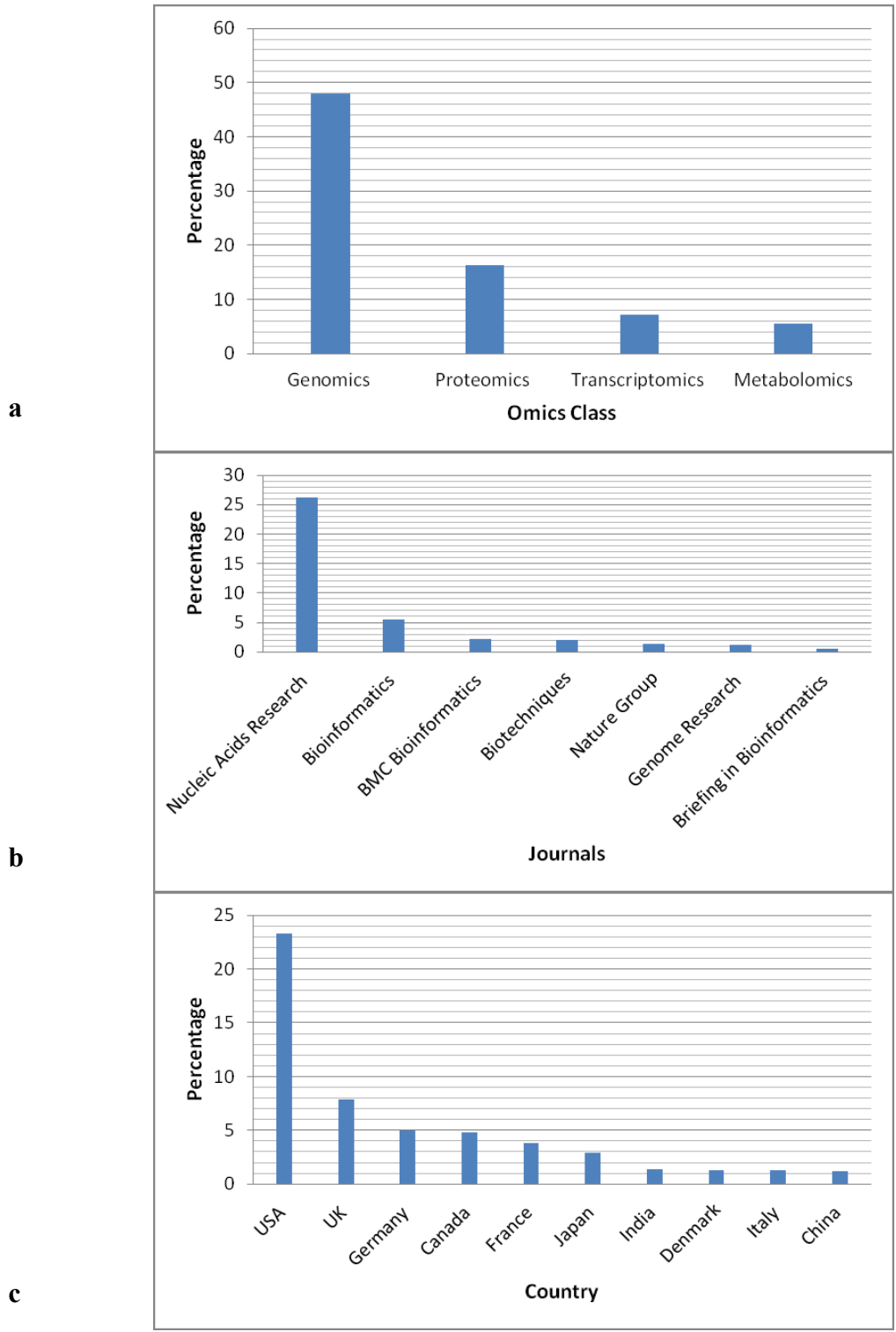

**Figure S1.** The barplot of the BioInfoBase records measured in terms of Omics classification (a) as well as Journals (b) and Countries (c) contributed in publishing and development of bioinformatics tools, databases and resources.

**Table 1.** List of the features extracted from the submitted records to produce the structured data tables as input to MySQL to generate searchable data tables by end user.

| Feature | Explanation | Example |
| --- | --- | --- |
| **Timestamp** | Date and time of the record registration (automatically inserted) | *2015-02-10 , 05:37:34* |
| **Original Code** | The record code provided by the admin | *G-S-SM-719* |
| **Record Creator** | Name of the person who provided the record | *Admin* |
| **Record Registrar** | Name of the person completing the record | *Dana Dereeper* |
| **Record Maintainer** | Name of the person who verifies and updates the record | *Dana Dereeper* |
| **Name** | Record name as the following format; Acronym of the record (if any) Full name of the record | *REDfly: Regulatory Element Database for Drosophila* |
| **First Category** | The main category to which the record belongs | *Database* |
| **Second Category** | One or more subcategories to which the above main category belongs | *Genomics* |
| **Application** | The general use of the record | *Molecular dynamics simulation* |
| **Organism Common Name** | Common name of the organism(s) which the record focuses on | *Rice, Mammalian* |
| **Organism Scientific Name** | Scientific name of the organism(s) which the record focuses on | *Oryza sativa* |
| **Authors/Developers** | Name of the person(s) who developed the software/database/… | *Kuo-Chen Chou, Gautam B. Singh, Hong-Bin Shen* |
| Keywords | Major keywords (3-6) related to the record as could be found in relevant article, if any. | *EST, expression, regulatory elements, rice, RT-PCR* |
| Features | Detailed information about the applications of the record | *- Four modes of manual alignment.*<br>*- In-color alignment and editing …* |
| Platform | Accessibility to the record (online/offline/both) | *Online* |

| Feature | Explanation | Example |
| --- | --- | --- |
| Operating System | System compatibility of the record | *Windows* |
| Other System Requirements | | *512 MB RAM required, 2 GB RAM recommended, 1024 x 768 display recommended, 32 and 64 bit platforms supported, Intel or AMD CPU required* |
| Programming Language | The programming language used to develop the record | *Python* |
| License | Availability of the record as free or commercial | *Free* |
| Price | Price of the facilities provided by the record | *1500 USD* |
| Version | Updated version of the record | *6.0.5* |
| Abstract | Abstract of the relevant article, if any. | *PLACE is a database of nucleotide sequence motifs found in plant cis-acting regulatory DNA elements. …* |
| Publication Citation | Citation of the relevant article (first and last version) according to NAR (Nucleic Acids Research) style | *Dong,X., Stothard,P., Forsythe,I.J. and Wishart,D.S. (2004) PlasMapper: a web server for drawing and auto-annotating plasmid maps. Nucleic Acids Res., 32, W660–W664.* |
| Journal | The journal that published the relevant article, if any | *Nucleic Acids Research* |
| Journal Impact Factor (IF) | Impact factor of the journal that published the article under citation, if any | *8.278* |
| Publisher | Name of the publisher to which the journal belongs, if any | *Oxford University Press* |
| Google Scholar Citations | The number of citations of the relevant article given by Google Scholar, if any | *Cited by 934* |
| Web Link | The link of record's home page | *http://www.dna.affrc.go.jp/PLACE/index.html* |
| Contact Information | Postal address of the record's developer, if any | *Bioinformatics Center, Institute for Chemical Research, Kyoto University, Uji, Kyoto 611-0011, Japan* |
| Email | Email address of the record developer | *xxx@ibc.wustl.edu* |

| Feature | Explanation | Example |
|---|---|---|
| **Institution** | Name of the institute where the record was developed | *Institute for Biomedical Computing, Washington University* |
| **Country** | Name the country where the institute is located | *USA* |
| **User ranking** | Usefulness of the record (ranking from 1-5) based on the registrar's experience | *3* |
| **Submit Review** | Advantages or disadvantages of the record based on the registrar's experience. | *MacVector is a comprehensive application that provides …* |
| **Tutorial** | Web link for the record tutorials/webinars, if any | *http://geneious.com/tutorials* |
| **Article Link** | Web link for the record article, if any | *http://david.abcc.ncifcrf.gov/forum/viewtopic.php?f…* |
| **Miscellaneous** | Any additional information that could be extracted from the record. | |

**Table 2.** A brief overview on the world bioinformatics resources landscape according to the submitted records. Numbers show the percentage of contribution in development and/or publication of bioinformatics resources based on the BioInfoBase database (Beta version).

| Omics Categories | | Leading Journals | | Leading Institutions | Leading Countries | |
|---|---|---|---|---|---|---|
| Genomics | 48.04 | Nucleic Acids Research | 26.2 | European Molecular Biology Laboratory-European Bioinformatics Institute (EMBL-EBI) | USA | 23.28 |
| Proteomics | 16.24 | Bioinformatics | 5.52 | | UK | 7.88 |
| Transcriptomics | 7.2 | BMC Bioinformatics | 2.12 | Agricultural-Food and Nutritional Science (AFNS), Bioinformatics and Genomics Department | Germany | 4.96 |
| Metabolomics | 5.52 | Biotechniques | 2.08 | | Canada | 4.76 |
| | | Nature Group | 1.4 | National Center for Biotechnology Information (NCBI) | Japan | 2.92 |
| | | Genome Research | 1.24 | | France | 3.84 |
| | | Briefing in Bioinformatics | 0.48 | National Institute of Health (NIH) | India | 1.4 |
| | | | | Swiss Bioinformatics Institute (SIB) | Denmark | 1.32 |
| | | | | Institut national de la recherche agronomique (INRA) | Italy | 1.32 |
| | | | | | China | 1.2 |

**References**

1. Gilbert,D. (2004) Bioinformatics software resources. *Brief. Bioinform.*, **5**, 300–304.

2. Brazas,M.D., Yim,D., Yeung,W. and Ouellette,B.F.F. (2012) A decade of Web Server updates at the Bioinformatics Links Directory: 2003-2012. *Nucleic Acids Res.*, **40**, W3–W12.

3. Brazas,M.D., Yim,D.S., Yamada,J.T. and Ouellette,B.F.F. (2011) The 2011 Bioinformatics Links Directory update: more resources, tools and databases and features to empower the bioinformatics community. *Nucleic Acids Res.*, **39**, W3–7.

4. Fox,J. a, Butland,S.L., McMillan,S., Campbell,G. and Ouellette,B.F.F. (2005) The Bioinformatics Links Directory: a compilation of molecular biology web servers. *Nucleic Acids Res.*, **33**, W3–24.

5. Bioinformatics Links Directory (2014).

6. Hsls.pitt.edu, (2015) OBRC: Online Bioinformatics Resources Collection | HSLS.

7. Shodhaka.com/startbioinfo, (2014) Startbioinfo. (2014).

8. Weblab.cbi.pku.edu.cn, (2015) WebLab.

9. Softberry.com, (2015) Softberry Home Page.

10. Omictools.com, (2015) A workflow for multi-omic data analysis - OMICtools.

11. Elements.eaglegenomics.com, (2015) The Elements of Bioinformatics.

12. Evolution.genetics.washington.edu, (2015) Phylogeny Programs.

13. Cederholm,D. (2010) CSS3 for web designers A Book Apart, New York.

14. Crowther,R. (2013) Hello! HTML5 & CSS3 a user-friendly reference guide Manning, Shelter Island, N.Y.

15. Safronov,M. (2014) Web Application Development with Yii 2 and PHP Packt Publishing, City.

16. Rose,S., Engel,D., Cramer,N. and Cowley,W. (2010) Automatic keyword extraction from individual documents. *Text Min.*

17. Salton,G. and Buckley,C. (1988) Term-weighting approaches in automatic text retrieval. *Inf. Process. Manag.*, **24**, 513–523.

18. Indurkhya,N. and Damerau,F.J. (2010) Handbook of natural language processing CRC Press.

19. Berry,M.W. and Kogan,J. (2010) Text mining: applications and theory John Wiley \& Sons.